\documentclass[epj-spec]{svjour}

\usepackage{amsmath,amssymb,bbm}

\newcommand{\assign}{:=}
\newcommand{\bigintlim}{\int}
\newcommand{\bignone}{\,}

\newcommand{\mathd}{\mathrm{d}}
\newcommand{\mathe}{\mathrm{e}}
\newcommand{\mathi}{\mathrm{i}}
\newcommand{\tmem}[1]{{\em #1\/}}
\newcommand{\tmmathbf}[1]{\ensuremath{\boldsymbol{#1}}}
\newcommand{\tmop}[1]{\ensuremath{\operatorname{#1}}}

\def\epjrunnhead{\markboth{}{}}%
\let\ProcessRunnHead=\epjrunnhead
\def\makeheadbox{\hfill\textsf{To appear in Eur. Phys. J. Special
    Topics (2007)}}

\begin{document}

\author{Bassano Vacchini\inst{1}\fnmsep\thanks{www.mi.infn.it/\~{}vacchini} 
\and
Klaus Hornberger\inst{2}\fnmsep\thanks{www.klaus-hornberger.de}
}
\institute{Dipartimento di Fisica
 dell'Universit\`a di Milano and INFN
 Sezione di Milano,\\ Via Celoria 16, 20133 Milano, Italy
\and Arnold Sommerfeld Center for Theoretical Physics,
Ludwig-Maximilians-Universit\"at M\"unchen,\\ Theresienstraße 37, 80333 Munich, Germany
}


\title{Relaxation dynamics of a quantum {Brownian} particle in an ideal
gas}

\abstract{
  We show how the quantum analog of the Fokker-Planck equation for describing
  Brownian motion can be obtained as the diffusive limit of the quantum linear
  Boltzmann equation. The latter describes the quantum dynamics of a tracer
  particle in a dilute, ideal gas by means of a translation-covariant master
  equation. We discuss the type of approximations required to obtain the
  generalized form of the Caldeira-Leggett master equation, along with their
  physical justification. Microscopic expressions for the diffusion and
  relaxation coefficients are obtained by analyzing the limiting form of the
  equation in both the Schr\"odinger and the Heisenberg picture.
}\maketitle

\section{Introduction}

\subsection{Quantum Brownian Motion}

One of the classic problems in open quantum dynamics is the question of
quantum Brownian motion, asking how a distinguished `Brownian' quantum
particle experiences friction, diffusion, and thermalization due to the
interaction with an unobserved surrounding liquid or gas. Starting with the
work of Caldeira and Leggett {\cite{Caldeira1983a}} the bulk of studies on
this problem treat the environment in a phenomenological way, usually by
linearly coupling the Brownian particle position to a continuous thermal bath
of harmonic oscillators, whose spectral density is then chosen as to reproduce
the desired relaxation and diffusion constants.

Using the Feynman-Vernon path integral approach {\cite{Feynman1963a}} these
linear models can even be solved exactly for some cases
{\cite{Haake1985a,Grabert1988a,Unruh1989a,Hu1992a}}. However, as is well
known, these non-Markovian dynamical solutions also have limitations. Firstly,
they usually have to assume that the Brownian particle and the environment are
initially in a product state, leading to an unphysical initial transient
dynamics due to the re-adjustment of the energies once the coupling is
switched on {\cite{Tameshtit1996a}}. Secondly, the generic assumption of a
linear coupling with the unbounded position operator, leading to spatial
correlations over any length scale, can be justified at best for a restricted
class of initial states. It will be valid if the Brownian particle state is
close to a classical state, but may lead to unphysically large decoherence
rates if the Brownian state is characterized by macroscopically large
coherence lengths {\cite{Hornberger2007LNP}}.

In a more concise sense, one should therefore characterize those situations as
generic quantum Brownian motion ({\tmem{i}}) where one is interested in
timescales such that the Markov assumption is permissible and ({\tmem{ii}})
where the motional states considered are close to a classical state in the
sense that the coherence scales are not macroscopic. This is the regime of the
{\tmem{Caldeira-Leggett master equation}} for free quantum Brownian motion
{\cite{Caldeira1983a}}. It is obtained from the path integral approach in a
high-temperature limit and, apart from the temperature $T$, it contains a
friction constant $\eta$ as a phenomenological parameter, see
Sect.~\ref{sec:ops} below. As in the corresponding case of the classical
Kramers equation (Sect.~\ref{sec:ps}), the diffusion constant $D_{pp}$ is
determined by $\eta$ and $T$ according to $D_{pp} = \eta Mk_{\text{B}} T$,
with $M$ the mass of the Brownian particle. This is an instance of the
fluctuation-dissipation theorem.

An issue of much debate is the fact that the Caldeira-Leggett master equation
is not of Lindblad-form and therefore does not preserve the positivity of some
initial states. As is well known {\cite{Breuer2002a}}, this can be healed by
adding a ``position-diffusion'' term provided the corresponding coefficient
$D_{xx}$ satisfies $D_{xx} \geqslant \eta^2 \hbar^2 / \left( 16 D_{pp}
\right)$, which in turn leads to ``momentum localization''. Various authors
proposed the `correct' value for $D_{xx}$ based either on formal arguments, on
improved evaluations of the high-temperature limit, and, above all, on more
specific descriptions of the environmental interaction process
{\cite{Tameshtit1996a,Breuer2002a,Sandulescu1987a,Diosi1993a,Diosi1995a,Halliwell1995a,Barnett2005a}}
. In view of this, it seems quite plausible that a definite answer concerning
the value of $D_{xx}$ can only be given by accounting for the environment in a
way which is microscopically more realistic.

\subsection{Microscopically realistic environments}

Given the large body of work on quantum Brownian motion it is perhaps
surprising that attempts to justify the Caldeira-Leggett master equation by a
realistic microscopic description of the environment received much less
attention. The main reason is the notorious difficulty of obtaining realistic
effective dynamic equations for the Brownian quantum particle in terms of the
microscopic properties of a given liquid or gas. Provided such a realistic
master equation is available in Lindblad form, one has then to take an
appropriate limit, which will be called ``{\tmem{diffusive limit}}'' below, to
end up with a master equation of the form of Caldeira-Leggett plus the
``position-diffusion'' term. Importantly, the coefficients $\eta$, $D_{pp}$,
and $D_{xx}$ are then no longer phenomenological constants, but they are
determined by the microscopic description of the particular environment
considered.

The simplest realistic environment in that sense is clearly given by an ideal
gas in a thermal state. The gas particles then do not interact with each
other, but they influence the Brownian particle via two-body forces, which
should be taken sufficiently short-ranged to permit a scattering theory
description of the interaction processes. The corresponding effective equation
of motion for the quantum Brownian particle is called {\tmem{quantum linear
Boltzmann equation}}{\footnote{This is the quantum analogue of the classical
linear Boltzmann equation for a tracer particle {\cite{Cercignani1975a}}. It
is important not to confuse this equation, which is non-perturbative, with the
{\tmem{linearized}} quantum equation for the reduced single particle gas state
of a self-interacting quantum gas.

}}. Such a Lindblad master equation was obtained recently by implementing the
Markov assumption before performing the partial trace over the gas particles
in a non-perturbative calculation {\cite{Vacchini2000a,Hornberger2006b}}. It
is supposed to be valid if the gas is sufficiently dilute to justify both the
neglect of three-body collisions and the Markov approximation. Moreover,
unlike the master equations for quantum Brownian motion, it is supposed to be
valid even for very non-classical motional states, such as the superposition
states found in an interferometer.

The purpose of the present article is to use this quantum linear Boltzmann
equation as a starting point for obtaining the extended Caldeira-Leggett
master equation. We will discuss what kind of assumptions and approximations
are required to end up in that form and how this {\tmem{diffusive limit}} can
be justified by using both the operator and the Wigner-Weyl formulation of
quantum mechanics in the Schr\"odinger and the Heisenberg picture. In all
cases we will find that the friction and diffusion coefficients $\eta$ and
$D_{pp}$ are uniquely specified by combining the thermodynamic quantities of
the gas with the relevant microscopic properties of its constituent particles,
namely their mass and their suitably averaged cross section. Moreover, we will
see that the coefficient of the ``position-diffusion'' term $D_{xx}$, which is
required for complete positivity, is completely determined by $\eta$ and
$D_{pp}$ at its smallest possible value, $D_{xx} = \eta^2 \hbar^2 / \left( 16
D_{pp} \right)$.

The structure of the article is as follows. In Sect.~\ref{sec:QLBE} we
briefly review the form of the quantum linear Boltzmann equation in operator
representation and in the momentum basis. Section \ref{sec:Schrodinger}
discusses the diffusive limit of the equation in the Schr\"odinger picture,
both in operator form and in the Wigner-Weyl phase space representation. In
Sect.~\ref{sec:Heisenberg} we formulate the same limit in the Heisenberg
picture and discuss the equations of motion for the energy and momentum
operator. We present our conclusions in Sect. \ref{sec:cc}.

\begin{flushleft}
  \section{The quantum linear Boltzmann equation}\label{sec:QLBE}
\end{flushleft}

Before we present the full form of the quantum linear Boltzmann equation let
us briefly collect some important steps that lead to its derivation. The idea
of using a scattering theory formulation for obtaining Markovian master
equations goes back to the work by Joos and Zeh on collisional decoherence
{\cite{Joos1985a}}. However, their master equation, which was later formulated
in a non-perturbative fashion {\cite{Gallis1990a,Hornberger2003b}}, cannot
describe friction, since the Brownian mass $M$ is assumed to be infinitely
large compared to the gas mass $m$, so that energy exchange cannot be
accommodated. An early proposal for finite mass ratios $m / M$ is the master
equation of Di\'osi {\cite{Diosi1995a}}, which is quite close to the present
formulation of the quantum linear Boltzmann equation, but, as we will see, it
differs in some crucial aspects, such as the inferred value of $D_{xx}$. A
perturbative form of the present quantum linear Boltzmann equation was
obtained in {\cite{Vacchini2000a,Vacchini2001a,Vacchini2001b}}, pointing to a
connection with the van Hove relation and the dynamic structure factor of the
gas. While other important contributions dealt with specific aspects
{\cite{Dodd2003a}}, the present non-perturbative form is a quite recent result
{\cite{Hornberger2006b}}, based on a monitoring approach for deriving
Markovian master equations {\cite{Hornberger2007b}}.
\vspace*{\baselineskip}

\subsection{Operator form}

In the following, we will denote the density operator for the motional state
of the Brownian particle by $\rho$, and its momentum operator by $\mathsf{P}$.
The quantum linear Boltzmann equation then reads
\begin{eqnarray}
  \frac{\mathd}{\mathd t} \rho & = & \frac{1}{\mathi \hbar} \left[
  \frac{\mathsf{P}^2}{2 M}, \rho \right] + \mathcal{L} \rho,  \label{eq:me0}
\end{eqnarray}
where the mapping $\mathcal{L}$ which describes the incoherent effects of the
gas environment is given by {\cite{Hornberger2006b}}
\begin{eqnarray}
  \mathcal{L} \rho & = & \bigintlim \mathd \tmmathbf{Q}
  \int_{\tmmathbf{Q}^{\bot}} \mathd \tmmathbf{p} \left\{ \mathsf{L}
  _{\tmmathbf{Q}, \tmmathbf{p}} \rho \mathsf{L} _{\tmmathbf{Q},
  \tmmathbf{p}}^{\dag} - \frac{1}{2} \rho \mathsf{L} _{\tmmathbf{Q},
  \tmmathbf{p}}^{\dag} \mathsf{L} _{\tmmathbf{Q}, \tmmathbf{p}} - \frac{1}{2} 
  \mathsf{L} _{\tmmathbf{Q}, \tmmathbf{p}}^{\dag} \mathsf{L} _{\tmmathbf{Q},
  \tmmathbf{p}} \rho \right\} \bignone .  \label{qlbe} \label{eq:qlbe}
\end{eqnarray}
Here the integration is over all momentum transfers $\tmmathbf{Q}$, and for
fixed $\tmmathbf{Q}$ also over the perpendicular plane $\tmmathbf{Q}^{\bot} =
\left\{ \tmmathbf{p} \in \mathbbm{R}^3 : \tmmathbf{p} \cdot \tmmathbf{Q}= 0
\right\}$. The Lindblad operators have the form
\begin{eqnarray}
  \mathsf{L} _{\tmmathbf{Q}, \tmmathbf{p}} & = & \mathe^{i\tmmathbf{Q} \cdot
  \mathsf{X} / \hbar} L \left( \tmmathbf{p}, \mathsf{P} ; \tmmathbf{Q} \right)
  \label{eq:Ldef}
\end{eqnarray}
where $\mathsf{X}$ is the position operator of the Brownian particle. The
first term in (\ref{eq:Ldef}) thus effects a momentum transfer determined by
$\tmmathbf{Q}$. It is important to stress that the form (\ref{eq:qlbe}) of
$\mathcal{L} \rho$ fits the general structure of a translation-covariant and
completely positive master equation as characterized by Holevo
{\cite{Holevo1996a}} (see {\cite{Petruccione2005a,Vacchini2005a}} for a
discussion).

The function $L$, which is operator-valued in (\ref{eq:Ldef}), contains all
the details of the collisional interaction with the gas. It involves the
momentum distribution function $\mu \left( \tmmathbf{p} \right)$ of the gas,
its number density $n_{\tmop{gas}}$ and the elastic scattering amplitude $f
\left( \tmmathbf{p}_f, \tmmathbf{p}_i \right)$, which determines the
differential cross section
\begin{eqnarray}
  \sigma \left( \tmmathbf{p}_f, \tmmathbf{p}_i \right) & = & \left| f \left(
  \tmmathbf{p}_f, \tmmathbf{p}_i \right) \right|^2, \nonumber
\end{eqnarray}
as well as the total cross section
\begin{eqnarray}
  \sigma_{\tmop{tot}} \left( \tmmathbf{p}_i \right) & = & \int \mathd
  \tmmathbf{n} \bignone \left| f \left( p_i \tmmathbf{n}, \tmmathbf{p}_i
  \right) \right|^2, \nonumber
\end{eqnarray}
where $\tmmathbf{n}$ is a unit vector with $\mathd \tmmathbf{n}$ the
associated solid angle element.

In order to specify $L$ let us denote, for any given momentum exchange
$\tmmathbf{Q} \neq 0$, the parallel and the perpendicular contribution of a
vector $\tmmathbf{P}$ by $\tmmathbf{P}_{\|\tmmathbf{Q}} = \left( \tmmathbf{P}
\cdot \tmmathbf{Q} \right) \tmmathbf{Q}/ Q^2$ and by $\tmmathbf{P}_{\bot
\tmmathbf{Q}} =\tmmathbf{P}-\tmmathbf{P}_{\|\tmmathbf{Q}}$, respectively. With
these definitions the function $L$ is defined by
\begin{eqnarray}
  L \left( \tmmathbf{p}, \tmmathbf{P}; \tmmathbf{Q} \right) & = &
  \sqrt{\frac{n_{\tmop{gas}} m}{Qm_{\ast}^2 }} \mu^{\frac{1}{2}} \left(
  \tmmathbf{p}_{\bot \tmmathbf{Q}} + \left( 1 + \frac{m}{M} \right)
  \frac{\tmmathbf{Q}}{2} + \frac{m}{M} \tmmathbf{P}_{\|\tmmathbf{Q}}
  \right)^{} \nonumber\\
  &  & \times f \left( \tmop{rel} \left( \tmmathbf{p}_{\bot \tmmathbf{Q}},
  \tmmathbf{P}_{\bot \tmmathbf{Q}} \right) - \frac{\tmmathbf{Q}}{2},
  \tmop{rel} \left( \tmmathbf{p}_{\bot \tmmathbf{Q}}, \tmmathbf{P}_{\bot
  \tmmathbf{Q}} \right) + \frac{\tmmathbf{Q}}{2} \right) .  \label{eq:Fdef}
\end{eqnarray}
The most natural choice for $\mu$ is of course the Maxwell-Boltzmann
distribution
\begin{eqnarray}
  \mu_{\beta} \left( \tmmathbf{p} \right) & = & \frac{1}{\pi^{3 / 2}
  p_{\beta}^3} \exp \left( - \frac{\tmmathbf{p}^2}{p_{\beta}^2} \right) 
  \label{eq:muMB}
\end{eqnarray}
with $p_{\beta}^2 = 2 m / \beta$ the most probable momentum at temperature $T
= 1 / \left( k_{\text{B}} \beta \right)$. Moreover, in (\ref{eq:Fdef}) we have
denoted the reduced mass by $m_{\ast} = mM / \left( M + m \right)$ and
relative momenta by
\begin{eqnarray*}
  \tmop{rel} \left( \tmmathbf{p}, \tmmathbf{P} \right) & \assign &
  \frac{m_{\ast}}{m} \tmmathbf{p}- \frac{m_{\ast}}{M} \tmmathbf{P}.
\end{eqnarray*}

One limiting form of the quantum linear Boltzmann equation is obtained by
replacing the scattering amplitudes in (\ref{eq:Fdef}) by their Born
approximation. This simplifies the equation considerably, since the Born
amplitude depends only on the difference of the momenta, $f_B \left(
\tmmathbf{p}_f, \tmmathbf{p}_i \right) = f_B \left( \tmmathbf{p}_f
-\tmmathbf{p}_i, 0 \right)$, which removes the operator-valuedness of the
scattering amplitudes. In this approximation one has
\begin{eqnarray}
  L_B \left( \tmmathbf{p}, \tmmathbf{P}; \tmmathbf{Q} \right) & = &
  \sqrt{\frac{n_{\tmop{gas}} m}{Qm_{\ast}^2 }} \mu^{\frac{1}{2}} \left(
  \tmmathbf{p}_{\bot \tmmathbf{Q}} + \left( 1 + \frac{m}{M} \right)
  \frac{\tmmathbf{Q}}{2} + \frac{m}{M} \tmmathbf{P}_{\|\tmmathbf{Q}} \right)
  f_B \left( -\tmmathbf{Q}, 0 \right) \nonumber
\end{eqnarray}
so that the $\mathd \tmmathbf{p}_{\bot \tmmathbf{Q}}$-integration in
(\ref{eq:qlbe}) can be carried out. The resulting equation thus reduces to the
one proposed in {\cite{Vacchini2000a,Vacchini2001a}}.

Another limiting form of the master equation is the case of an infinitely
massive Brownian particle, $m / M \rightarrow 0$, where it describes no
dissipation, but pure spatial decoherence. As one expects, the quantum linear
Boltzmann equation reduces in this limit to the proper master equation for
collisional decoherence
{\cite{Hornberger2003b,Hornberger2004a,Vacchini2005a}}, which was recently
tested experimentally {\cite{Hornberger2003a,Hackermuller2003b,Arndt2005a}}.

\subsection{Momentum representation}

Although the operator form (\ref{eq:qlbe})-(\ref{eq:muMB}) of $\mathcal{L}$
will be quite useful for the discussion of the diffusive limit below, the
physics described by the master equation is more easily understood in momentum
representation. Let us denote matrix elements of $\rho$ in the basis of
improper momentum eigenkets as $\langle \tmmathbf{P}| \rho |\tmmathbf{P}'
\rangle = \rho \left( \tmmathbf{P}, \tmmathbf{P}' \right)$. The incoherent
part (\ref{eq:qlbe}) of the quantum linear Boltzmann equation then takes the
form \
\begin{eqnarray}
  \langle \tmmathbf{P}| \mathcal{L} \rho |\tmmathbf{P}' \rangle & \text{} = &
  \int \mathd \tmmathbf{Q} \bignone \rho \left( \tmmathbf{P}-\tmmathbf{Q},
  \tmmathbf{P}' -\tmmathbf{Q} \right) M_{\tmop{in}} \left( \tmmathbf{P},
  \tmmathbf{P}' ; \tmmathbf{Q} \right) \nonumber\\
  &  & - \rho \left( \tmmathbf{P}, \tmmathbf{P}' \right)  \frac{1}{2} \int
  \mathd \tmmathbf{Q} \left[ M_{\tmop{in}} \left( \tmmathbf{P}+\tmmathbf{Q},
  \tmmathbf{P}+\tmmathbf{Q}; \tmmathbf{Q} \right) + M_{\tmop{in}} \left(
  \tmmathbf{P}' +\tmmathbf{Q}, \tmmathbf{P}' +\tmmathbf{Q}; \tmmathbf{Q}
  \right) \right] \nonumber
\end{eqnarray}
with the complex function $M_{\tmop{in}} \left( \tmmathbf{P}, \tmmathbf{P}' ;
\tmmathbf{Q} \right)$ defined by
\begin{eqnarray}
  M_{\tmop{in}} \left( \tmmathbf{P}, \tmmathbf{P}' ; \tmmathbf{Q} \right) & =
  & \int_{\tmmathbf{Q}^{\bot}} \mathd \tmmathbf{p}L \left( \tmmathbf{p},
  \tmmathbf{P}-\tmmathbf{Q}; \tmmathbf{Q} \right) L^{\ast} \left(
  \tmmathbf{p}, \tmmathbf{P}' -\tmmathbf{Q}; \tmmathbf{Q} \right) . 
  \label{eq:Min3}
\end{eqnarray}
One can show that for $\tmmathbf{P}=\tmmathbf{P}'$ this function is equal to
the rate density, found in the classical linear Boltzmann equation, of the
particle ending up with momentum $\tmmathbf{P}$ after a momentum gain of
$\tmmathbf{Q}$ due to a gas collision. In other words, we have $M_{\tmop{in}}
\left( \tmmathbf{P}, \tmmathbf{P}; \tmmathbf{Q} \right) =
M_{\tmop{in}}^{\tmop{cl}} \left( \tmmathbf{P}; \tmmathbf{Q} \right)$ with
\begin{eqnarray}
  M^{\tmop{cl}}_{\tmop{in}} \left( \tmmathbf{P}; \tmmathbf{Q} \right) & =
  \bignone & \frac{n_{\tmop{gas}}}{m_{\ast}} \bigintlim \mathd \tmmathbf{p}_0
  \mu \left( \tmmathbf{p}_0 \right) \sigma \left( \tmop{rel} \left(
  \tmmathbf{p}_0 -\tmmathbf{Q}, \tmmathbf{P} \right), \tmop{rel} \left(
  \tmmathbf{p}_0, \tmmathbf{P}-\tmmathbf{Q} \right) \right) \nonumber\\
  &  & \phantom{\frac{n}{m_{\ast}^2} \bigintlim} \times \delta \left(
  \frac{\left| \tmop{rel} \left( \tmmathbf{p}_0 -\tmmathbf{Q}, \tmmathbf{P}
  \right) \right|^2 - \left| \tmop{rel} \left( \tmmathbf{p}_0,
  \tmmathbf{P}-\tmmathbf{Q} \right) \right|^2}{2} \right) .  \label{eq:Mclin}
\end{eqnarray}
We can equally consider the corresponding classical rate density for the
particle with momentum $\tmmathbf{P}$ to end up at a different momentum,
\begin{eqnarray}
  M_{\tmop{out}}^{\tmop{cl}} \left( \tmmathbf{P} \right) & \assign & \int
  \mathd \tmmathbf{Q} \, M_{\tmop{in}}^{\tmop{cl}} \left(
  \tmmathbf{P}+\tmmathbf{Q}; \tmmathbf{Q} \right) \nonumber\\
  & = & \frac{n_{\tmop{gas}}}{m_{\ast}} \bigintlim \mathd \tmmathbf{p}_0
  \mathd \tmmathbf{Q} \mu \left( \tmmathbf{p}_0 \right) \sigma \left(
  \tmop{rel} \left( \tmmathbf{p}_0, \tmmathbf{P} \right) +\tmmathbf{Q},
  \tmop{rel} \left( \tmmathbf{p}_0, \tmmathbf{P} \right) \right) \nonumber\\
  &  & \phantom{\frac{n}{m_{\ast}^2} \bigintlim} \times \delta \left(
  \frac{\left| \tmop{rel} \left( \tmmathbf{p}_0, \tmmathbf{P} \right)
  \right|^2 - \left| \tmop{rel} \left( \tmmathbf{p}_0, \tmmathbf{P} \right)
  +\tmmathbf{Q} \right|^2}{2} \right) .  \label{eq:Mclout}
\end{eqnarray}
This can be used to put the master equation into the shorter form
\begin{eqnarray}
  \langle \tmmathbf{P}| \mathcal{L} \rho |\tmmathbf{P}' \rangle & = & \int
  \mathd \tmmathbf{Q} \bignone M_{\tmop{in}} \left( \tmmathbf{P},
  \tmmathbf{P}' ; \tmmathbf{Q} \right) \rho \left( \tmmathbf{P}-\tmmathbf{Q},
  \tmmathbf{P}' -\tmmathbf{Q} \right) \nonumber\\
  &  & - \frac{1}{2} \left[ M_{\tmop{out}}^{\tmop{cl}} \left( \tmmathbf{P}
  \right) + M_{\tmop{out}}^{\tmop{cl}} \left( \tmmathbf{P}' \right) \right]
  \rho \left( \tmmathbf{P}, \tmmathbf{P}' \right) .  \label{eq:Dtrho2}
\end{eqnarray}
For $\tmmathbf{P} \neq \tmmathbf{P}'$ the function $M_{\tmop{in}}$ is in
general complex and cannot be related to a classical rate, but still the
analogy to the classical case seems quite intuitive.

\section{Diffusive limit in the Schr\"odinger picture}\label{sec:Schrodinger}

We now want to consider the Brownian motion limit of the quantum linear
Boltzmann equation, using arguments similar to the treatment that turns the
classical linear Boltzmann equation into the Fokker-Planck equation
{\cite{Uhlenbeck1948a}}. The situation is actually more complicated in the
quantum case, since one is dealing with operators whose values can be
estimated in a meaningful way only when suitable matrix elements are
considered.

We will argue that the quantum counterpart of the classical Fokker-Planck
equation can be obtained by formally expanding the operators in the collision
kernel of the quantum linear Boltzmann equation up to second order
contributions in the canonically conjugate operators $\mathsf{X}$ and
$\mathsf{P}$. As we shall see, the result is {\tmem{not}} equivalent to a
naive application of the correspondence principle on the classical result.
Such a procedure would simply lead to the original Caldeira-Leggett master
equation {\cite{Caldeira1983a}}, which does not guarantee to preserve the
positivity of the statistical operator. The operator expansion holds under
conditions analogous to the classical ones {\cite{Uhlenbeck1948a}}, which
imply in particular that the statistical operator describes a very massive
test particle not far from thermal equilibrium, that is to say, close to
diagonal in momentum representation. More specifically, the off-diagonal
elements $\langle \tmmathbf{P}| \rho |\tmmathbf{P}' \rangle$ may differ
significantly from zero only for $\Delta P \assign \left|
\tmmathbf{P}-\tmmathbf{P}' \right| \lesssim \sqrt{M / \beta}$. In the position
representation the validity of the expansion requires that the statistical
operator is only coherent over a length of the order of the thermal wavelength
of the test particle, so that $\langle \tmmathbf{X}| \rho |\tmmathbf{X}'
\rangle$ is appreciably different from zero only within a range given by the
thermal de Broglie wave length $\lambda_{\tmop{th}} = \sqrt{2 \pi \hbar^2
\beta / M}$, i.e., for $\Delta X \assign \left| \tmmathbf{X}-\tmmathbf{X}'
\right| \lesssim \sqrt{2 \pi \hbar^2 \beta / M}$. Note that since we require
$m / M \ll 1$ it follows that the Brownian thermal wave length is much smaller
that the thermal wave length of the gas, $\lambda_{\tmop{th}} \ll
\lambda_{\tmop{th}}^{\tmop{gas}} = \sqrt{2 \pi \hbar^2 \beta / m}$.

\subsection{Operator formulation\label{sec:ops}}

In order to formulate the Brownian motion limit of the quantum linear
Boltzmann equation we come back to its explicit expression
(\ref{eq:qlbe})-(\ref{eq:Fdef}) and we confine ourselves to the case of a
{\tmem{constant}} scattering cross-section $\left| f \left( \tmmathbf{p}_f,
\tmmathbf{p}_i \right) \right|^2 = \sigma_{\tmop{tot}} / 4 \pi$. In this case,
the operator form of the quantum linear Boltzmann equation is given explicitly
by
\begin{eqnarray}
  \mathcal{L} \rho & = & n_{\tmop{gas}}  \frac{m}{m^2_{\ast}}
  \frac{\sigma_{\tmop{tot}}}{4 \pi} \bigintlim \frac{\mathd \tmmathbf{Q}}{Q}
  \int_{\tmmathbf{Q}^{\bot}} \mathd \tmmathbf{p}  \label{eq:full}\\
  &  & \times \left[ \mathe^{i\tmmathbf{Q} \cdot \mathsf{X} / \hbar}
  \mu^{\frac{1}{2}} \left( \tmmathbf{p}_{\bot \tmmathbf{Q}} \text{$+
  \frac{m}{m_{\ast}^{}} \frac{\tmmathbf{Q}}{2} + \frac{m}{M}
  \mathbf{\mathsf{P}}_{\|\tmmathbf{Q}}$} \right) \rho \mu^{\frac{1}{2}} \left(
  \tmmathbf{p}_{\bot \tmmathbf{Q}} \text{$+ \frac{m}{m_{\ast}^{}}
  \frac{\tmmathbf{Q}}{2} + \frac{m}{M} \mathbf{\mathsf{P}}_{\|\tmmathbf{Q}}$} \right)
  \mathe^{- i\tmmathbf{Q} \cdot \mathsf{X} / \hbar} \right. \nonumber\\
  &  & \left. - \frac{1}{2} \left\{ \mu \left( \tmmathbf{p}_{\bot
  \tmmathbf{Q}} \text{$+ \frac{m}{m_{\ast}^{}} \frac{\tmmathbf{Q}}{2} +
  \frac{m}{M} \mathbf{\mathsf{P}}_{\|\tmmathbf{Q}}$} \right), \rho \right\} \right]
  \nonumber
\end{eqnarray}
where $\left\{ \cdot, \cdot \right\}$ denotes the anti-commutator.
Specializing to the case of a Maxwell-Boltzmann distribution in the gas
(\ref{eq:muMB}) we can write
\begin{eqnarray}
  \mathcal{L} \rho & = & n_{\tmop{gas}}  \frac{m}{m^2_{\ast}}
  \frac{\sigma_{\tmop{tot}}}{4 \pi} \bigintlim \frac{\mathd \tmmathbf{Q}}{Q}
  \int_{\tmmathbf{Q}^{\bot}} \mathd \tmmathbf{p}\, \mu_{\beta} \left(
  \tmmathbf{p}_{\bot \tmmathbf{Q}} \text{$+ \frac{m}{m_{\ast}^{}}
  \frac{\tmmathbf{Q}}{2}$} \right)  \nonumber\\
  &  & \times \left[ \mathe^{i\tmmathbf{Q} \cdot \mathsf{X} / \hbar} \exp
  \left( - \beta \frac{m \mathsf{P^2_{\|\tmmathbf{Q}}}}{4 M^2} \mathsf{} -
  \beta \frac{m\tmmathbf{Q} \cdot \mathsf{P}}{4 Mm_{\ast}^{}} \right)^{} \rho
  \exp \left( - \beta \frac{m \mathsf{P^2_{\|\tmmathbf{Q}}}}{4 M^2} - \beta
  \frac{m\tmmathbf{Q} \cdot \mathsf{P}}{4 Mm_{\ast}^{}} \right)^{} \mathe^{-
  i\tmmathbf{Q} \cdot \mathsf{X} / \hbar} \right. \nonumber\\
  &  & \left.  \phantom{\times \left[ \right.} - \frac{1}{2} \left\{ \exp
  \left( - \beta \frac{m \mathsf{P^2_{\|\tmmathbf{Q}}}}{4 M^2} \mathsf{} -
  \beta \frac{m\tmmathbf{Q} \cdot \mathsf{P}}{4 Mm_{\ast}^{}} \right), \rho
  \right\} \right] . \nonumber
\end{eqnarray}
Since we are ultimately interested in an expansion up to second order in
$\mathsf{P}$, the contributions coming from the terms involving squares of the
momentum operator in the exponent will simply cancel out. We can replace them
by unity, leading to the much simpler expression
\begin{eqnarray}
  \mathcal{L} \rho & = & n_{\tmop{gas}}  \frac{m}{m^2_{\ast}}
  \frac{\sigma_{\tmop{tot}}}{4 \pi} \bigintlim \frac{\mathd \tmmathbf{Q}}{Q}
  \int_{\tmmathbf{Q}^{\bot}} \mathd \tmmathbf{p}\, \mu_{\beta} \left(
  \tmmathbf{p}_{\bot \tmmathbf{Q}} \text{$+ \frac{m}{m_{\ast}^{}}
  \frac{\tmmathbf{Q}}{2}$} \right)  \nonumber\\
  &  & \times \left[ \mathe^{i\tmmathbf{Q} \cdot \mathsf{X} / \hbar} \exp
  \left( - \beta \frac{m\tmmathbf{Q} \cdot \mathsf{P}}{4 Mm_{\ast}^{}} \right)
  \rho \exp \left( - \beta \frac{m\tmmathbf{Q} \cdot \mathsf{P}}{4
  Mm_{\ast}^{}} \right) \mathe^{- i\tmmathbf{Q} \cdot \mathsf{X} / \hbar} -
  \frac{1}{2} \left\{ \exp \left( - \beta \frac{m\tmmathbf{Q} \cdot
  \mathsf{P}}{2 Mm_{\ast}^{}} \right), \rho \right\} \right] . \nonumber
\end{eqnarray}
For a small mass ratio $m / M \ll 1$ this yields finally
\begin{eqnarray}
  \mathcal{L} \rho & = & \frac{n_{\tmop{gas}} }{m}
  \frac{\sigma_{\tmop{tot}}}{4 \pi} \bigintlim \frac{\mathd \tmmathbf{Q}}{Q}
  \int_{\tmmathbf{Q}^{\bot}} \mathd \tmmathbf{p}\, \mu_{\beta} \left(
  \tmmathbf{p}_{\bot \tmmathbf{Q}} \text{$+ \frac{\tmmathbf{Q}}{2}$} \right)  
  \label{eq:sensible}\\
  &  & \times \left[ \mathe^{i\tmmathbf{Q} \cdot \mathsf{X} / \hbar} \exp
  \left( - \beta \frac{\tmmathbf{Q} \cdot \mathsf{P}}{4 M} \right) \rho \exp
  \left( - \beta \frac{\tmmathbf{Q} \cdot \mathsf{P}}{4 M} \right) \mathe^{-
  i\tmmathbf{Q} \cdot \mathsf{X} / \hbar} - \frac{1}{2} \left\{ \exp \left( -
  \beta \frac{\tmmathbf{Q} \cdot \mathsf{P}}{2 M} \right), \rho \right\}
  \right], \nonumber
\end{eqnarray}
which we will use as the starting point for expanding in $\mathsf{X}$ and
$\mathsf{P}$. The main requirement for the expansion is the assumption that
the change in momentum of the Brownian particle is small compared to the
scales involved in its motional state. More specifically, one has to assume
that for typical values of the momentum transfer $Q$ the relevant matrix
elements of the statistical operator vanish unless
\begin{eqnarray}
  \frac{Q}{\hbar} \Delta X & \ll & 1 \nonumber
\end{eqnarray}
and
\begin{eqnarray}
  \frac{\beta Q}{M} \Delta P & \ll & 1. \nonumber
\end{eqnarray}
These conditions are both satisfied if the Brownian state is close to thermal
and $M \gg m$ since the momentum transfer $Q$ is then typically of the order
of the momentum of the colliding gas particles $p_{\beta} = \sqrt{2 m /
\beta}$. This implies in particular that
\begin{eqnarray}
  \frac{Q}{\hbar} \approx \sqrt{\frac{m}{2 \pi \hbar^2 \beta}} =
  \frac{1}{\lambda_{\tmop{th}}^{\tmop{gas}}} \ll \frac{1}{\lambda_{\tmop{th}}}
  \lesssim \frac{1}{\Delta X} &  &  \nonumber
\end{eqnarray}
and
\begin{eqnarray}
  \frac{\beta Q}{M} \approx \sqrt{\frac{m}{M}} \sqrt{\frac{\beta}{M}} \ll
  \sqrt{\frac{\beta}{M}} \lesssim \frac{1}{\Delta P} . &  &  \nonumber
\end{eqnarray}
Expanding the terms in the square brackets of (\ref{eq:sensible}) one thus
arrives at
\begin{eqnarray}
  - \frac{1}{2} \sum^3_{i, j = 1} Q_i Q_j \left[ \frac{1}{\hbar^2} \left[
  \mathsf{X}_i, \left[ \mathsf{X}_j, \rho \right] \right] + \left(
  \frac{\beta}{4 M} \right)^2 \left[ \mathsf{P}_i, \left[ \mathsf{P}_j, \rho
  \right] \right] + \frac{i}{\hbar} \frac{\beta}{2 M} \left[ \mathsf{X}_i,
  \left\{ \mathsf{P}_j, \rho \right\} \right] \right] \bignone, &  & 
  \nonumber
\end{eqnarray}
where a term linear in the momentum transfer has been omitted since it
vanishes upon the integration in (\ref{eq:sensible}). It follows that one has
to evaluate the integrals
\begin{eqnarray}
  \eta_{ij} & = & \frac{\beta}{2 M}  \frac{n_{\tmop{gas}} }{m}
  \frac{\sigma_{\tmop{tot}}}{4 \pi} \bigintlim \frac{\mathd \tmmathbf{Q}}{Q}
  \int_{\tmmathbf{Q}^{\bot}} \mathd \tmmathbf{p}\, \mu_{\beta} \left(
  \tmmathbf{p}_{\bot \tmmathbf{Q}} \text{$+ \frac{\tmmathbf{Q}}{2}$} \right)
  Q_i Q_j, \nonumber\\
  & = & \delta_{ij} \frac{\beta}{6 M} \frac{n_{\tmop{gas}} }{m}
  \frac{\sigma_{\tmop{tot}}}{4 \pi} \bigintlim_{} \mathd \tmmathbf{Q}Q \text{}
  \int_{\tmmathbf{Q}^{\bot}} \mathd \tmmathbf{p}\, \mu_{\beta} \left(
  \tmmathbf{p}_{\bot \tmmathbf{Q}} \text{$+ \frac{\tmmathbf{Q}}{2}$} \right)
  \hspace{0.6em} = \hspace{0.6em} \delta_{ij} \eta . \nonumber
\end{eqnarray}
The coefficient $\eta$ is given by
\begin{eqnarray}
  \eta & = & \frac{\beta}{6 M} \frac{n_{\tmop{gas}} }{m}
  \frac{\sigma_{\tmop{tot}}}{4 \pi} \bigintlim_{} \mathd \tmmathbf{Q}Q
  \mathe^{- \beta Q^2 / \left( 8 m \right)} \int_{\tmmathbf{Q}^{\bot}} \mathd
  \tmmathbf{p} \, \mu_{\beta} \left( \tmmathbf{p}_{\bot \tmmathbf{Q}} \text{}
  \right) \nonumber\\
  & = & \frac{16}{3} n_{\tmop{gas}} \sigma_{\tmop{tot}}
  \sqrt{\frac{mk_{\text{B}} T}{2 \pi M^2} } .  \label{eq:eta}
\end{eqnarray}
The final result thus takes the form
\begin{eqnarray}
  \mathcal{L} \rho & = & - \frac{i}{\hbar} \frac{\eta}{2} \sum^3_{i = 1}
  \left[ \mathsf{X}_i, \left\{ \mathsf{P}_i, \rho \right\} \right] -
  \frac{D_{pp}}{\hbar^2} \sum^3_{i = 1} \left[ \mathsf{X}_i, \left[
  \mathsf{X}_i, \rho \right] \right] - \frac{D_{xx}}{\hbar^2} \sum^3_{i = 1}
  \left[ \mathsf{P}_i, \left[ \mathsf{P}_i, \rho \right] \right], 
  \label{eq:me}
\end{eqnarray}
which is an extended version of the Caldeira-Leggett equation. In the original
master equation the third term is absent, $D_{xx} = 0$, and $\eta$ is a
phenomenological parameter. In contrast, the diffusive limit of the quantum
linear Boltzmann equation yielded a microscopically defined friction constant
(\ref{eq:eta}), and also the diffusion coefficients $D_{pp}$ and $D_{xx}$ are
directly related to $\eta$,
\begin{eqnarray}
  D_{pp} & = & \eta Mk_{\text{B}} T  \label{eq:Dppfinal}
\end{eqnarray}
and
\begin{eqnarray}
  D_{xx} & = & \eta \frac{\hbar^2}{16 Mk_B T} \hspace{0.1em} = \hspace{0.1em}
  \left( \frac{\hbar^{}}{4 Mk_{\text{B}} T} \right)^2 D_{pp} . 
  \label{eq:Dxxfinal}
\end{eqnarray}
Equation (\ref{eq:Dppfinal}) is the expected expression of the
fluctuation-dissipation relation, while the coefficient (\ref{eq:Dxxfinal}) of
the ``position-diffusion term'' $\sum_i \bignone [ \mathsf{P}_i, [
\mathsf{P}_i, \rho]]$ has just the minimal value required to ensure the
preservation of the positivity of the statistical operator with elapsing time
{\cite{Breuer2002a}}. Since this new contribution appears only in the quantum
case, it cannot be read out from the classical Fokker-Planck equation and had
to be fixed on the basis of a microscopic derivation at the quantum level, as
pointed out in {\cite{Vacchini2000a}}. Note in particular, that Di\'osi's form
of the linear quantum Boltzmann equation {\cite{Diosi1995a}} leads to a
different expression of the coefficient $D_{xx}$, which is not just a function
of $\eta$ and T.

Concerning the comparison with the classical Brownian motion, the considered
case of a constant scattering cross-section applies in the classical
formulation to perfectly rigid spheres {\cite{Uhlenbeck1948a}}, and in that
case the relevant classical scattering cross-section is the geometric one,
$\sigma^{\tmop{cl}}_{\tmop{tot}} = \pi R^2$, with $R$ radius of the sphere. We
have just shown that the same friction coefficient appears for a constant
scattering cross-section in the quantum case provided $R \gg
\lambda_{\tmop{th}}^{\tmop{gas}}$, due to the fact that the forward scattering
contribution then cancels out in the master-equation
{\cite{Schiff1968a,Adler2006a}}. To see how the particular choice
(\ref{eq:Dxxfinal}) of $D_{xx}$ relates to the classical description it is now
helpful to consider Eq. (\ref{eq:me}) in the Wigner-Weyl phase space
formulation.

\subsection{Phase space description\label{sec:ps}}

In view of the appearance of a quantum mechanically required
``position-diffusion'' term in the expression (\ref{eq:me}) it is of interest
to follow the transition from quantum linear Boltzmann equation to quantum
Fokker-Planck by means of the Wigner function {\cite{Vacchini2002a}}, which
despite the fact that it is not a proper probability density allows for a
classical phase-space picture of the quantum dynamics. As usual, we denote the
Wigner function associated to a statistical operator $\rho$ as
\begin{eqnarray}
  W \left( \tmmathbf{X}, \tmmathbf{P} \right) & = & \int \frac{\mathd
  \tmmathbf{K}}{\left( 2 \pi \hbar^{} \right)^3} \mathe^{i\tmmathbf{X} \cdot
  \tmmathbf{K}/ \hbar} \langle \tmmathbf{P}+ \frac{\tmmathbf{K}}{2} | \rho
  |\tmmathbf{P}- \frac{\tmmathbf{K}}{2} \rangle \bignone .  \label{eq:Wigner}
\end{eqnarray}
Equation (\ref{eq:sensible}) then reads
\begin{eqnarray}
  \frac{\partial}{\partial t} W \left( \tmmathbf{X}, \tmmathbf{P} \right) & =
  & n_{\tmop{gas}}  \frac{\sigma_{\tmop{tot}}}{4 \pi} \frac{m}{m^2_{\ast}}
  \bigintlim \frac{\mathd \tmmathbf{Q}}{Q} \int_{\tmmathbf{Q}^{\bot}} \mathd
  \tmmathbf{p} \, \mu_{\beta} \left( \tmmathbf{p}_{\bot \tmmathbf{Q}} \text{$+
  \frac{\tmmathbf{Q}}{2}$} \right)  \int \frac{\mathd \tmmathbf{K}}{\left( 2
  \pi \hbar^{} \right)^3} \mathe^{i\tmmathbf{X} \cdot \tmmathbf{K}/ \hbar}
  \nonumber\\
  &  & \times \left[ \exp \left( - \beta \frac{\tmmathbf{Q} \cdot \left(
  \tmmathbf{P}-\tmmathbf{Q} \right)}{2 M} \right) \langle
  \tmmathbf{P}-\tmmathbf{Q}+ \frac{\tmmathbf{K}}{2} | \rho
  |\tmmathbf{P}-\tmmathbf{Q}- \frac{\tmmathbf{K}}{2} \rangle \right.
  \nonumber\\
  &  & \phantom{\times \left[ \right.} - \exp \left( - \beta
  \frac{\tmmathbf{Q} \cdot \left( \tmmathbf{P}+\tmmathbf{K}/ 2 \right)}{2 M}
  \right) \langle \tmmathbf{P}+ \frac{\tmmathbf{K}}{2} | \rho |\tmmathbf{P}-
  \frac{\tmmathbf{K}}{2} \rangle \nonumber\\
  &  & \left. \phantom{\times \left[ \right.} - \exp \left( - \beta
  \frac{\tmmathbf{Q} \cdot \left( \tmmathbf{P}-\tmmathbf{K}/ 2 \right)}{2 M}
  \right) \langle \tmmathbf{P}+ \frac{\tmmathbf{K}}{2} | \rho |\tmmathbf{P}-
  \frac{\tmmathbf{K}}{2} \rangle \right] . \nonumber
\end{eqnarray}
A closed equation for $W\left( \tmmathbf{X}, \tmmathbf{P} \right)$ is obtained by inserting the inverse of
(\ref{eq:Wigner}),
\begin{eqnarray}
  \frac{\partial}{\partial t} W \left( \tmmathbf{X}, \tmmathbf{P} \right) & =
  & n_{\tmop{gas}}  \frac{\sigma_{\tmop{tot}}}{4 \pi} \frac{m}{m^2_{\ast}}
  \bigintlim \frac{\mathd \tmmathbf{Q}}{Q} \int_{\tmmathbf{Q}^{\bot}} \mathd
  \tmmathbf{p} \, \mu_{\beta} \left( \tmmathbf{p}_{\bot \tmmathbf{Q}} \text{$+
  \frac{\tmmathbf{Q}}{2}$} \right)  \nonumber\\
  &  & \times \left[ \exp \left( - \beta \frac{\tmmathbf{Q} \cdot \left(
  \tmmathbf{P}-\tmmathbf{Q} \right)}{2 M} \right) W \left( \tmmathbf{X},
  \tmmathbf{P}-\tmmathbf{Q} \right) \right. \nonumber\\
  &  & \left. \phantom{\times \left[ \right.} - \exp \left( - \beta
  \frac{\tmmathbf{Q} \cdot \tmmathbf{P}}{2 M} \right) \cosh \left( \frac{\beta
  \hbar}{4 M} \tmmathbf{Q} \cdot \nabla_{\tmmathbf{X}} \right) W \left(
  \tmmathbf{X}, \tmmathbf{P} \right) \right] \nonumber\\
  & = & n_{\tmop{gas}}  \frac{\sigma_{\tmop{tot}}}{4 \pi}
  \frac{m}{m^2_{\ast}} \bigintlim \frac{\mathd \tmmathbf{Q}}{Q}
  \int_{\tmmathbf{Q}^{\bot}} \mathd \tmmathbf{p}\, \mu_{\beta} \left(
  \tmmathbf{p}_{\bot \tmmathbf{Q}} \text{$+ \frac{\tmmathbf{Q}}{2}$} \right) 
  \nonumber\\
  &  & \times \left[ \exp \left( -\tmmathbf{Q} \cdot \nabla_{\tmmathbf{P}}
  \right) - \cosh \left( \frac{\beta \hbar}{4 M} \tmmathbf{Q} \cdot
  \nabla_{\tmmathbf{X}} \right) \right] \exp \left( - \beta \frac{\tmmathbf{Q}
  \cdot \tmmathbf{P}}{2 M} \right) W \left( \tmmathbf{X}, \tmmathbf{P} \right)
  . \nonumber
\end{eqnarray}
where $\cosh$ denotes the hyperbolic cosine. The equation can now be written
more compactly as
\begin{eqnarray}
  \frac{\partial}{\partial t} W \left( \tmmathbf{X}, \tmmathbf{P} \right) & =
  & n_{\tmop{gas}}  \frac{\sigma_{\tmop{tot}}}{4 \pi} \frac{m}{m^2_{\ast}}
  \bigintlim \frac{\mathd \tmmathbf{Q}}{Q} \int_{\tmmathbf{Q}^{\bot}} \mathd
  \tmmathbf{p} \, \mu_{\beta} \left( \tmmathbf{p}_{\bot \tmmathbf{Q}} \text{$+
  \frac{\tmmathbf{Q}}{2}$} \right)  \label{eq:wigner}\\
  &  & \times \left[ \exp \left( -\tmmathbf{Q} \cdot \nabla_{\tmmathbf{P}}
  \right) - \cosh \left( \frac{\beta \hbar}{4 M} \tmmathbf{Q} \cdot
  \nabla_{\tmmathbf{X}} \right) \right] \exp \left( - \beta \frac{\tmmathbf{Q}
  \cdot \tmmathbf{P}}{2 M} \right) W \left( \tmmathbf{X}, \tmmathbf{P} \right)
  \nonumber
\end{eqnarray}
where we have used the unitary differential operators $\exp \left(
-\tmmathbf{Q} \cdot \nabla_{\tmmathbf{P}} \right)$ and ${\exp (\beta \hbar /
\left( 4 M \right) \tmmathbf{Q} \cdot \nabla_{\tmmathbf{X}})}^{}$ effecting a
shift in momentum and position, respectively, of the arguments of the Wigner
function. As one can check, Eq. (\ref{eq:wigner}) differs from the
corresponding classical expression of the linear Boltzmann equation
{\cite{Williams1966}} just by the hyperbolic cosine term, which in the
classical case is replaced by unity, as one would obtain in the naive
classical limit $\hbar \rightarrow 0$. An expansion up to second order of the
exponential operators appearing in the kernel of (\ref{eq:wigner}) transforms
the integro-differential equation for the Wigner function into a partial
differential equation. It is the phase space representation of Eq.
(\ref{eq:me}) and reads,
\begin{eqnarray}
  \frac{\partial}{\partial t} W \left( \tmmathbf{X}, \tmmathbf{P} \right) & =
  & \eta \nabla_{\tmmathbf{P}} \cdot \left( \tmmathbf{P}W \left( \tmmathbf{X},
  \tmmathbf{P} \right) \right) + D_{pp} \Delta_{\tmmathbf{P}} W \left(
  \tmmathbf{X}, \tmmathbf{P} \right) + D_{xx} \Delta_{\tmmathbf{X}} W \left(
  \tmmathbf{X}, \tmmathbf{P} \right) . \nonumber
\end{eqnarray}
This is the quantum counterpart of the classical Fokker-Planck equation, for
the time evolution of the probability density $f_{\tmop{cl}} \left(
\tmmathbf{X}, \tmmathbf{P} \right)$ in phase-space,
\begin{eqnarray}
  \frac{\partial}{\partial t} f_{\tmop{cl}} \left( \tmmathbf{X}, \tmmathbf{P}
  \right) & = & \eta \nabla_{\tmmathbf{P}} \cdot \left(
  \tmmathbf{P}f_{\tmop{cl}} \left( \tmmathbf{X}, \tmmathbf{P} \right) \right)
  + D_{pp} \Delta_{\tmmathbf{P}} f_{\tmop{cl}} \left( \tmmathbf{X},
  \tmmathbf{P} \right) . \nonumber
\end{eqnarray}
The quantum and the classical phase space equations for Brownian motion thus
differ again by the appearance of a ``position-diffusion'' term, which is
symmetric with respect to the regular diffusion term.

\section{Diffusive limit in the Heisenberg picture}\label{sec:Heisenberg}

In the previous paragraphs we understood the quantum linear Boltzmann equation
as a mapping $\mathcal{L}$ acting on the statistical operator, thus working in
the Schr\"odinger picture. In the same spirit we considered its diffusive
limit, which led to the quantum counterpart of the classical Fokker-Planck
equation. In the following we will consider its adjoint mapping
$\mathcal{L}^{\ast}$ for the time evolution of observables. It is defined
through the relation
\begin{eqnarray}
  \tmop{Tr} \left( \mathsf{A} \mathcal{L} \rho \right) & = & \tmop{Tr} \left(
  \rho \mathcal{L}^{\ast} \mathsf{A} \right), \nonumber
\end{eqnarray}
where the trace operation expresses the duality relation between the space of
trace class operators, which contains the states given by statistical
operators $\rho$, and its dual, the space of bounded operators $\mathsf{A}$
characterizing observables.

Using the form (\ref{eq:qlbe})-(\ref{eq:Ldef}) of $\mathcal{L}$ this leads to
the explicit identification
\begin{eqnarray}
  \mathcal{L}^{\ast} \mathsf{A} & = & \bigintlim \mathd \tmmathbf{Q}
  \int_{\tmmathbf{Q}^{\bot}} \mathd \tmmathbf{p} \left[ L^{\dagger} \left(
  \tmmathbf{p}, \mathsf{P} ; \tmmathbf{Q} \right) \mathe^{- i\tmmathbf{Q}
  \cdot \mathsf{X} / \hbar} \mathsf{A} \mathe^{i\tmmathbf{Q} \cdot \mathsf{X}
  / \hbar} L \left( \tmmathbf{p}, \mathsf{P} ; \tmmathbf{Q} \right)  
  \label{eq:Hqlbe} \right.\\
  &  & \left. - \frac{1}{2} \left\{ L^{\dagger} \left( \tmmathbf{p},
  \mathsf{P} ; \tmmathbf{Q} \right) L \left( \tmmathbf{p}, \mathsf{P} ;
  \tmmathbf{Q} \right), \mathsf{A} \right\} \right] \bignone . \nonumber
\end{eqnarray}
The full differential equation for the time evolution of an Heisenberg
operator $\mathsf{A}_t$ is thus given by
\begin{eqnarray}
  \frac{\mathd}{\mathd t} \mathsf{A}_t & = & \frac{1}{\mathi \hbar} \left[
  \mathsf{A}_t, \frac{\mathsf{P}^2}{2 M} \right] + \int \bignone \mathd
  \tmmathbf{Q} \int_{\tmmathbf{Q}^{\bot}} \mathd \tmmathbf{p} \left[
  L^{\dagger} \left( \tmmathbf{p}, \mathsf{P} ; \tmmathbf{Q} \right) \mathe^{-
  i\tmmathbf{Q} \cdot \mathsf{X} / \hbar} \mathsf{A}_t \mathe^{i\tmmathbf{Q}
  \cdot \mathsf{X} / \hbar} L \left( \tmmathbf{p}, \mathsf{P} ; \tmmathbf{Q}
  \right)  \right. \nonumber\\
  &  & \left. - \frac{1}{2} \left\{ L^{\dagger} \left( \tmmathbf{p},
  \mathsf{P} ; \tmmathbf{Q} \right) L \left( \tmmathbf{p}, \mathsf{P} ;
  \tmmathbf{Q} \right), \mathsf{A}_t \right\} \right] \bignone . \nonumber
\end{eqnarray}
We will take the Heisenberg operator to coincide with the corresponding
Schr\"odinger observable at $t = 0$, i.e., $\mathsf{A}_0 = \mathsf{A}$.

An important class of Schr\"odinger picture observables are those which are
only functions of the momentum operator, $\mathsf{A} = A \left( \mathsf{P}
\right)$. The equation of motion in the Heisenberg picture then simplifies
considerably,
\begin{eqnarray}
  \frac{\mathd}{\mathd t} \mathsf{A}_t & = & \int \bignone \mathd \tmmathbf{Q}
  \int_{\tmmathbf{Q}^{\bot}} \mathd \tmmathbf{p} \left| L \left( \tmmathbf{p},
  \mathsf{P} ; \tmmathbf{Q} \right) \right|^2 \left[  \text{$\mathe^{-
  i\tmmathbf{Q} \cdot \mathsf{X} / \hbar} \mathsf{A}_t \mathe^{i\tmmathbf{Q}
  \cdot \mathsf{X} / \hbar}$} - \mathsf{A}_t \right] . \nonumber
\end{eqnarray}
Note that $\mathcal{L}$, and equivalently $\mathcal{L}^{\ast}$, is covariant
under translations {\cite{Vacchini2001b}} in the sense that
\begin{eqnarray}
  \mathcal{L} \left[ \mathe^{- i\tmmathbf{b} \cdot \mathsf{\mathsf{P}} /
  \hbar} \rho \mathe^{i\tmmathbf{b} \cdot \mathsf{\mathsf{P}} / \hbar} \right]
  & = & \mathe^{- i\tmmathbf{b} \cdot \mathsf{\mathsf{P}} / \hbar} \mathcal{L}
  \left[ \rho \right] \mathe^{i\tmmathbf{b} \cdot \mathsf{\mathsf{P}} / \hbar}
  . \nonumber
\end{eqnarray}
It follows that the algebra generated by the momentum operator is left
invariant. Consequently, $\mathsf{A}_0 = A_0 \left( \mathsf{P} \right)$
implies $\mathsf{A}_t = A_t ( \mathsf{P})$ for $t > 0$ and therefore $[
\mathsf{A}_t, \mathsf{P}] = 0$. In particular, recalling the definition of the
quantum and the classical in-rates in (\ref{eq:Min3}) and (\ref{eq:Mclin}),
respectively, we find that observables given by a function of momentum obey
\begin{eqnarray}
  \frac{\mathd}{\mathd t} \text{$A_t \left( \mathsf{P} \right)$} & = &
  \bignone \int \bignone \mathd \tmmathbf{Q}M^{\tmop{cl}}_{\tmop{in}} \left(
  \mathsf{P} +\tmmathbf{Q}; \tmmathbf{Q} \right) \left[ \text{$\mathe^{-
  i\tmmathbf{Q} \cdot \mathsf{X} / \hbar} A_t \left( \mathsf{P} \right)
  \mathe^{i\tmmathbf{Q} \cdot \mathsf{X} / \hbar}$} - A_t \left( \mathsf{P}
  \right) \right] \nonumber\\
  & = & \bignone \int \bignone \mathd \tmmathbf{Q}M^{\tmop{cl}}_{\tmop{in}}
  \left( \mathsf{P} +\tmmathbf{Q}; \tmmathbf{Q} \right) \left[ A_t \left(
  \mathsf{P} +\tmmathbf{Q} \right) - A_t \left( \mathsf{P} \right) \right],
  \nonumber
\end{eqnarray}
in strict analogy with the classical formulation.

We now focus on the time evolution of the expectation values of momentum and
kinetic energy, $A \left( \mathsf{P} \right) = \mathsf{P}$ and $A \left(
\mathsf{P} \right) = \mathsf{P}^2 / \left( 2 M \right)$, respectively. We
shall first obtain the explicit dynamic equations and then the limiting form
corresponding to the diffusive limit. Starting from the expression for the
time evolution of expectation values,
\begin{eqnarray}
  \frac{\mathd}{\mathd t} \langle \mathsf{A}_{} \rangle_{\rho_t} : =
  \frac{\mathd}{\mathd t} \tmop{Tr} \left( \mathsf{A} \rho_t \right) =
  \tmop{Tr} \left( \mathsf{A} \mathcal{L} \rho_t \right) = \tmop{Tr} \left(
  \rho_t \mathcal{\mathcal{L}^{\ast}} \mathsf{A} \mathcal{} \right), &  & 
  \label{eq:11}
\end{eqnarray}
we can now exploit the fact that functions of the momentum operator are mapped
by $\mathcal{L}^{\ast}$ to functions of the momentum operator. Specifically,
it is convenient to first evaluate
\begin{eqnarray}
  \mathcal{L}^{\ast} \left[ A \left( \mathsf{P} \right) \right] & = & \int
  \mathd \tmmathbf{Q} \bignone M^{\tmop{cl}}_{\tmop{in}} \left(
  \mathsf{P}+\tmmathbf{Q}; \tmmathbf{Q} \right) \left[ A \left( \mathsf{P}
  +\tmmathbf{Q} \right) - A \left( \mathsf{P} \right) \right],  \label{eq:hh}
\end{eqnarray}
which can be easily dealt with as an equation for $\mathbbm{C}$-numbers, by
working in the momentum basis. Moreover, assuming that the scattering
cross-section is invariant under parity transformations the scattering rate
(\ref{eq:Mclin}) satisfies
\begin{eqnarray}
  M^{\tmop{cl}}_{\tmop{in}} \left( \tmmathbf{P}+\tmmathbf{Q}; \tmmathbf{P}
  \right)  & \equiv & M^{\tmop{cl}}_{\tmop{in}} \left( \tmmathbf{P}
  \rightarrow \tmmathbf{P}+\tmmathbf{Q} \right)   \label{eq:paritym}\\
  & = & M^{\tmop{cl}}_{\tmop{in}} \left( -\tmmathbf{P} \rightarrow
  -\tmmathbf{P}-\tmmathbf{Q} \right) \nonumber\\
  & \equiv & M^{\tmop{cl}}_{\tmop{in}} \left( -\tmmathbf{P}-\tmmathbf{Q};
  -\tmmathbf{Q} \right), \nonumber
\end{eqnarray}
so that the operator $\mathcal{L}^{\ast} \left[ A \left( \mathsf{P} \right)
\right]$ has the same parity as $A \left( \mathsf{P} \right)$. According to
our convention momentum or energy transfers are positive when the test
particle gains momentum or energy. The fact that $M^{\tmop{cl}}_{\tmop{in}}$
is positive thus ensures the obvious physical requirement that the change of
momentum is positive when the momentum transfer is positive, and likewise that
a positive energy transfer increases the energy. In fact, for $A
(\tmmathbf{P}) =\tmmathbf{P}$ the quantity $\left[ A \left(
\tmmathbf{P}+\tmmathbf{Q} \right) - A \left( \tmmathbf{P} \right) \right]$ is
simply the momentum transfer in the single collision, while for $A
(\tmmathbf{P}) = P^2 / \left( 2 M \right)$ the quantity $\left[ A \left(
\tmmathbf{P}+\tmmathbf{Q} \right) - A \left( \tmmathbf{P} \right) \right]$ is
the energy transfer $E \left( \tmmathbf{Q}, \tmmathbf{P} \right) = \left(
\tmmathbf{P}+\tmmathbf{Q} \right)^2 / \left( 2 M \right) - P^2 / \left( 2 M
\right)$ in a single collision with momentum gain $\tmmathbf{Q}$.

Let us first write the explicit expression for $M^{\tmop{cl}}_{\tmop{in}}
\left( \tmmathbf{P}+\tmmathbf{Q}; \tmmathbf{P} \right)$ for a gas described by
the Maxwell-Boltzmann distribution (\ref{eq:muMB}):
\begin{eqnarray}
  M^{\tmop{cl}}_{\tmop{in}} \left( \tmmathbf{P}+\tmmathbf{Q}; \tmmathbf{Q}
  \right) & = & \frac{n_{\tmop{gas}} m}{m^2_{\ast} Q}
  \int_{\tmmathbf{Q}^{\bot}} \mathd \tmmathbf{p}\, \mu_{\beta} \left(
  \tmmathbf{p}_{\bot \tmmathbf{Q}} + \frac{m}{m_{\ast}} \frac{\tmmathbf{Q}}{2}
  + \frac{m}{M} \tmmathbf{P}_{\|\tmmathbf{Q}} \right) \nonumber\\
  &  & \times \sigma \left( \tmop{rel} \left( \mathbf{\tmmathbf{p}}_{\perp
  \tmmathbf{Q}}, \tmmathbf{P}_{\perp \tmmathbf{Q}} \right) -
  \frac{\tmmathbf{Q}}{2}, \tmop{rel} \left( \mathbf{\tmmathbf{p}}_{\perp
  \tmmathbf{Q}}, \tmmathbf{P}_{\perp \tmmathbf{Q}} \right) +
  \frac{\tmmathbf{Q}}{2} \right) .  \label{eq:51}
\end{eqnarray}
Inserting this into Eq. (\ref{eq:hh}) invites the more compact notation
\begin{eqnarray}
  \tilde{\sigma} \left( \tmmathbf{P}_{\perp \tmmathbf{Q}}, \tmmathbf{Q}
  \right) & \equiv & \int_{\tmmathbf{Q}^{\bot}} \mathd \tmmathbf{p}
  \, \mu_{\beta} \left( \tmmathbf{p}_{\bot \tmmathbf{Q}} \right) \sigma \left(
  \tmop{rel} \left( \mathbf{\tmmathbf{p}}_{\perp \tmmathbf{Q}},
  \tmmathbf{P}_{\perp \tmmathbf{Q}} \right) - \frac{\tmmathbf{Q}}{2},
  \tmop{rel} \left( \mathbf{\tmmathbf{p}}_{\perp \tmmathbf{Q}},
  \tmmathbf{P}_{\perp \tmmathbf{Q}} \right) + \frac{\tmmathbf{Q}}{2} \right), 
  \label{eq:tilda}
\end{eqnarray}
so that we can write
\begin{eqnarray}
  \mathcal{L}^{\ast} \left[ A \left( \mathsf{P} \right) \right] & = &
  \frac{n_{\tmop{gas}}}{m_{\ast}^2}  \sqrt{\frac{\beta m}{2 \pi}} \int
  \frac{\mathd \tmmathbf{Q}}{Q}  \tilde{\sigma} ( \mathsf{P}_{\perp},
  \tmmathbf{Q}) \exp \left( - \beta \frac{mQ^2}{8 m_{\ast}^2} - \beta \frac{m
  \mathsf{P}_{\|\tmmathbf{Q}}^2}{2 M^2} - \beta \frac{m\tmmathbf{Q} \cdot
  \text{$\mathsf{P}_{}$} }{2 Mm_{\ast}^{}} \right) \nonumber\\
  &  & \times \text{$\left[ A \left( \mathsf{P} +\tmmathbf{Q} \right) - A
  \left( \mathsf{P} \right) \right]$} .  \label{eq:hpicture}
\end{eqnarray}
We now proceed to evaluate this expression explicitly for the case of the
momentum and the kinetic energy observable, by considering the special case of
a constant scattering cross-section, $\left| f \left( \tmmathbf{p}_f,
\tmmathbf{p}_i \right) \right|^2 = \sigma_{\tmop{tot}} / 4 \pi$, as in Sect.
\ref{sec:Schrodinger}. In the momentum case, $A \left( \mathsf{P} \right) =
\mathsf{P}$, it is advantageous to rescale by the reference value
$Mv_{\beta}$, where $v_{\beta} = p_{\beta} / m = \sqrt{2 / \left( \beta m
\right)}$ is the most probable velocity of the gas particles.
\begin{eqnarray}
  \mathcal{L}^{\ast} \left[ \frac{\mathsf{P}}{Mv_{\beta}} \right] & = &
  \frac{n_{\tmop{gas}}}{\sqrt{\pi} m_{\ast}^2 v^2_{\beta} M} 
  \frac{\sigma_{\tmop{tot}}}{4 \pi} \int \mathd \tmmathbf{Q}
  \frac{\tmmathbf{Q}}{Q} \exp \left( - \beta \frac{mQ^2}{8 m_{\ast}^2} - \beta
  \frac{m \mathsf{P}_{\|\tmmathbf{Q}}^2}{2 M^2} - \beta \frac{m\tmmathbf{Q}
  \cdot \text{$\mathsf{P}_{}$} }{2 Mm_{\ast}^{}} \right) \nonumber\\
  & \equiv & I_1  \label{eq:pp}
\end{eqnarray}
Similarly, for the kinetic energy $A \left( \mathsf{P} \right) =
\mathsf{P^{}}^2 / \left( 2 M \right)$ the expression (\ref{eq:hpicture}) leads
to
\begin{eqnarray}
  \mathcal{L}^{\ast} \left[ \frac{1}{2 M} \left( \frac{\mathsf{P}}{Mv_{\beta}}
  \right)^2 \right] & = & \frac{n_{\tmop{gas}}}{\sqrt{\pi} m_{\ast}^2
  v^3_{\beta} M^2}  \frac{\sigma_{\tmop{tot}}}{4 \pi} \int \frac{\mathd
  \tmmathbf{Q}}{Q} \exp \left( - \beta \frac{mQ^2}{8 m_{\ast}^2} - \beta
  \frac{m \mathsf{P}_{\|\tmmathbf{Q}}^2}{2 M^2} - \beta \frac{m\tmmathbf{Q}
  \cdot \text{$\mathsf{P}_{}$} }{2 Mm_{\ast}^{}} \right) \nonumber\\
  &  & \times \left[ \frac{Q^2}{2 M} + \frac{\mathsf{P} \cdot
  \tmmathbf{Q}}{M} \right] \nonumber\\
  & \equiv & I_2 .  \label{eq:ee}
\end{eqnarray}
The evaluation of $I_1$ and $I_2$ is most easily done by focussing on their
functional expression, thus determining them as functions of the dimensionless
variable
\begin{eqnarray}
  \tmmathbf{U}= \frac{\tmmathbf{P}}{Mv_{\beta}} =
  \frac{\tmmathbf{V}}{v_{\beta}}, &  &  \nonumber
\end{eqnarray}
and carrying out the integration over the scaled momentum transfer
$\tmmathbf{K}$
\begin{eqnarray*}
  \tmmathbf{K}= \frac{\tmmathbf{Q}}{m_{\ast}^{} v_{\beta}} . &  & 
\end{eqnarray*}
>From the right hand side of (\ref{eq:pp}) we have
\begin{eqnarray}
  I_1 \left( \tmmathbf{U} \right) & = & n_{\tmop{gas}} 
  \frac{\sigma_{\tmop{tot}}}{4 \pi} \frac{v_{\beta}}{\sqrt{\pi}}
  \frac{m_{\ast}^{}}{M} \int \mathd \tmmathbf{K} \frac{\tmmathbf{K}}{K} \exp
  \left( - K^2  \left( \frac{1}{2} + \frac{\tmmathbf{U} \cdot
  \tmmathbf{K}}{K^2} \right)^2 \right),  \label{eq:66}
\end{eqnarray}
and writing
\begin{eqnarray}
  \tmmathbf{K} & = & \tmmathbf{K}_{\parallel \tmmathbf{U}} +\tmmathbf{K}_{\bot
  \tmmathbf{U}} \nonumber
\end{eqnarray}
one finds that only $\tmmathbf{K}_{\parallel \tmmathbf{U}}$, i.e., the
component parallel to $\tmmathbf{U}$, contributes to (\ref{eq:66}). We are
thus left with
\begin{eqnarray}
  I_1 \left( \tmmathbf{U} \right) & = & n_{\tmop{gas}} 
  \frac{\sigma_{\tmop{tot}}}{4 \pi} \frac{v_{\beta}}{\sqrt{\pi}}
  \frac{m_{\ast}^{}}{M} \int \mathd \tmmathbf{K} \frac{\tmmathbf{K}_{\parallel
  \tmmathbf{U}}}{K} \exp \left( - K^2  \left( \frac{1}{2} + \frac{\tmmathbf{U}
  \cdot \tmmathbf{K}}{K^2} \right)^2 \right) . \nonumber
\end{eqnarray}
Taking now the direction of $\tmmathbf{U}$ as the polar axis and denoting by
$\xi$ the cosine of the angle between $\tmmathbf{K}$ and $\tmmathbf{U}$ we
have to evaluate
\begin{eqnarray}
  I_1 \left( \tmmathbf{U} \right) & = & n_{\tmop{gas}} 
  \frac{\sigma_{\tmop{tot}}}{4 \pi} \frac{v_{\beta}}{\sqrt{\pi}}
  \frac{m_{\ast}^{}}{M} 2 \pi \frac{\tmmathbf{U}}{U} \int^{+ \infty}_0 \mathd
  K K^2  \int^{+ 1}_{- 1} \mathd \xi \, \xi \mathe^{- \left( K / 2 + U \xi
  \right)^2}  \label{eq:67}\\
  & = & - n_{\tmop{gas}}  \frac{\sigma_{\tmop{tot}}}{4 \pi} \sqrt{\frac{8
  \pi^{}}{m \beta}} \frac{_{} m_{\ast}^{}}{M}  \frac{\tmmathbf{U}}{U} \int^{+
  \infty}_0 \mathd^{} KK^2 \mathe^{- K^2 / 4} \nonumber\\
  &  & \times \left\{ \frac{\sqrt{\pi} K}{4 U^2} \mathe^{K^2 / 4} \left[
  \tmop{erf} \left( \frac{K}{2} + U \right) - \tmop{erf} \left( \frac{K}{2} -
  U \right) \right] - \frac{\mathe^{- U^2}}{U^2} \sinh (UK) \right\}
  \nonumber\\
  & = & - n_{\tmop{gas}}  \frac{\sigma_{\tmop{tot}}}{4 \pi} \sqrt{\frac{8
  \pi^{}}{m \beta}} \frac{_{} m_{\ast}^{}}{M}  \frac{\tmmathbf{U}}{U^2}
  \left\{ \left[ 1 + 2 U^2 \right] \mathe^{- U^2} - \left[ 1 - 4 U^2 - 4 U^4
  \right] \frac{\sqrt{\pi}}{2} \frac{\text{erf} (U)}{U}  \right\} . \nonumber
\end{eqnarray}
Here, $\text{erf} (x) = 2 \pi^{- \frac{1}{2}} \int^x_0 \exp \left( - t^2
\right) \mathd t$ denotes the error function. The expression (\ref{eq:67}) can
be further simplified and expressed more compactly by means of the confluent
hypergeometric function $\text{}_1 F_1$, in particular by using the known
expression of $\text{}_1 F_1 \left( \alpha, \gamma ; z \right)$ for indexes
$\alpha = - 1 / 2$ and $\gamma = 5 / 2$ {\cite{Wolfram}},
\begin{eqnarray}
  \text{}_1 F_1 \left( - \frac{1}{2}, \frac{5}{2} ; - U^2 \right) & = &
  \frac{3}{16} \frac{1}{U^2} \left\{ \left[ 1 + 2 U^2 \right] \mathe^{- U^2} -
  \left[ 1 - 4 U^2 - 4 U^4 \right] \frac{\sqrt{\pi}}{2} \frac{\text{erf}
  (U)}{U}  \right\} .  \label{eq:g3}
\end{eqnarray}
This way (\ref{eq:67}) finally becomes
\begin{eqnarray}
  I_1 \left( \tmmathbf{U} \right) & = & - n_{\tmop{gas}} 
  \frac{\sigma_{\tmop{tot}}}{4 \pi} \frac{16}{3} \sqrt{\frac{8 \pi^{}}{m
  \beta}} \frac{_{} m_{\ast}^{}}{M} \tmmathbf{U} \text{}_1 F_1 \left( -
  \frac{1}{2}, \frac{5}{2} ; - U^2 \right),  \label{eq:lastm}
\end{eqnarray}
with $\text{}_1 F_1 \left( - \frac{1}{2}, \frac{5}{2} ; - U^2 \right)$ a
positive, monotonically increasing function.

For the rescaled kinetic energy one has to consider the r.h.s. of
(\ref{eq:ee}), so that the function $I_2$ is given by
\begin{eqnarray}
  I_2 \left( U^2 \right) & = & 2 n_{\tmop{gas}}  \frac{\sigma_{\tmop{tot}}}{4
  \pi} \frac{v_{\beta}}{\sqrt{\pi}} \frac{m^{}_{\ast}}{M^{}} \int \mathd
  \tmmathbf{K} \exp \left( - K^2  \left( \frac{1}{2} + \frac{\tmmathbf{U}
  \cdot \tmmathbf{K}}{K^2} \right)^2 \right) \left[ \frac{1}{2}
  \frac{m^{}_{\ast}}{M^{}} K^{} + \frac{\tmmathbf{U} \cdot \tmmathbf{K}}{K}
  \right] . \nonumber
\end{eqnarray}
As before, we take the direction of $\tmmathbf{U}$ as polar axis and denote by
$\xi$ the cosine of the angle between $\tmmathbf{K}$ and $\tmmathbf{U}$. This
way,
\begin{eqnarray}
  I_2 \left( U^2 \right) & = & 2 n_{\tmop{gas}}  \frac{\sigma_{\tmop{tot}}}{4
  \pi} \sqrt{\frac{8 \pi^{}}{m \beta}} \frac{_{} m_{\ast}^{}}{M}  \int^{+
  \infty}_0 \mathd^{} KK^2 \int^{+ 1}_{- 1} \mathd \xi \mathe^{- \left( K / 2
  + U \xi \right)^2} \text{$\left[ \frac{1}{2} \frac{m^{}_{\ast}}{M^{}} K^{} +
  U \xi \right]$} \nonumber\\
  & = & - 2 n_{\tmop{gas}}  \frac{\sigma_{\tmop{tot}}}{4 \pi} \sqrt{\frac{8
  \pi^{}}{m \beta}} \frac{_{} m_{\ast}^{}}{M} \int^{+ \infty}_0 \mathd^{} KK^2
  \mathe^{- K^2 / 4} \nonumber\\
  &  & \times \left\{ \left( 1 - \frac{m^{}_{\ast}}{M^{}} \right)
  \frac{\sqrt{\pi} K}{4 U^{}} \mathe^{K^2 / 4} \left[ \tmop{erf} \left(
  \frac{K}{2} + U \right) - \tmop{erf} \left( \frac{K}{2} - U \right) \right]
  - \frac{\mathe^{- U^2}}{U^{}} \sinh (UK) \right\} \nonumber\\
  & = & - 2 n_{\tmop{gas}}  \frac{\sigma_{\tmop{tot}}}{4 \pi} \sqrt{\frac{8
  \pi}{m \beta}} \frac{_{} m_{\ast}^{}}{M} \left\{ \left[ 1 + 2 U^2 \right]
  \mathe^{- U^2} - \left[ 1 - 4 U^4 - 4 U^2 \right]  \frac{\sqrt{\pi}}{2} 
  \frac{\text{erf} (U)}{U} \right. \nonumber\\
  &  & \left. - \frac{_{} m_{\ast}^{}}{M} \left(  \left[ 5 + 2 U^2 \right]
  \mathe^{- U^2} + \left[ 3 + 12 U^2 + 4 U^4 \right] \frac{\sqrt{\pi}}{2} 
  \frac{\text{erf} \left( U \right)}{U}   \right)\right\} . \nonumber
\end{eqnarray}
Also this expression can be written much more compactly by exploiting again
(\ref{eq:g3}), as well as the known expression of $\text{}_1 F_1 \left( -
\frac{1}{2}, \frac{3}{2} ; - U^2 \right)$, as given in {\cite{Wolfram}}, and
the functional relations between confluent hypergeometric functions with
different indexes {\cite{Gradshteyn1965a}}. This yields
\begin{eqnarray}
  \text{}_1 F_1 \left( - \frac{3}{2}, \frac{3}{2} ; - U^2 \right) & = &
  \frac{1}{8} \left\{  \left[ 5 + 2 U^2 \right] \mathe^{- U^2} + \left[ 3 + 12
  U^2 + 4 U^4 \right] \frac{\sqrt{\pi}}{2}  \frac{\text{erf} \left( U
  \right)}{U}  \right\},  \label{eq:g4}
\end{eqnarray}
and therefore
\begin{eqnarray}
  I_2 \left( U^2 \right) & = & - 2 n_{\tmop{gas}} \frac{\sigma_{\tmop{tot}}}{4
  \pi}  \frac{16}{3}  \sqrt{\frac{8 \pi}{m \beta}} \frac{_{} m_{\ast}^{}}{M}
  \left\{ U^2 \text{}_1 F_1 \left( - \frac{1}{2}, \frac{5}{2} ; - U^2 \right) -
  \frac{3}{2} \frac{_{} m_{\ast}^{}}{M} \text{}_1 F_1 \left( - \frac{3}{2},
  \frac{3}{2} ; - U^2 \right) \right\}, \nonumber\\
  &  &  \label{eq:laste}
\end{eqnarray}
with both $\text{}_1 F_1 \left( - \frac{1}{2}, \frac{5}{2} ; - U^2 \right)$
and $\text{}_1 F_1 \left( - \frac{3}{2}, \frac{3}{2} ; - U^2 \right)$ positive
and monotonically increasing functions.

Finally, coming back to the operator expression with the original variables
(\ref{eq:pp})-(\ref{eq:ee}) the time evolution of the momentum expectation
values satisfies, according to (\ref{eq:11}),
\begin{eqnarray}
  \frac{\mathd}{\mathd t} \langle \mathsf{P} \rangle_{\rho_t} & = & -
  n_{\tmop{gas}}  \frac{\sigma_{\tmop{tot}}}{4 \pi} \frac{16}{3} \sqrt{\frac{8
  \pi^{}}{m \beta}} \frac{_{} m_{\ast}^{}}{M} \langle \mathsf{P}  \text{}_1
  F_1 \left( - \frac{1}{2}, \frac{5}{2} ; - \left( \frac{\mathsf{P}^{}}{M^{}
  v^{}_{\beta}} \right)^2 \right) \rangle_{\rho_t} .  \label{eq:p}
\end{eqnarray}
Similarly, we have for the kinetic energy $\mathsf{E} = \mathsf{P}^2 / \left(
2 M \right)$
\begin{eqnarray}
  \frac{\mathd}{\mathd t} \langle \mathsf{E} \rangle_{\rho_t} & = & - 2
  n_{\tmop{gas}} \frac{\sigma_{\tmop{tot}}}{4 \pi}  \frac{16}{3} 
  \sqrt{\frac{8 \pi}{m \beta}} \frac{_{} m_{\ast}^{}}{M}  \label{eq:e}\\
  &  & \times \langle \text{}_1 F_1 \left( - \frac{1}{2}, \frac{5}{2} ; -
  \beta \mathsf{E} \frac{m}{M} \right) \mathsf{E} - \frac{3}{2 \beta}
  \frac{_{} m_{\ast}^{} }{m}  \text{}_1 F_1 \left( - \frac{3}{2}, \frac{3}{2}
  ; - \beta \mathsf{E} \frac{m}{M} \right) \rangle_{\rho_t} . \nonumber
\end{eqnarray}
>From this it is immediately clear that in general there is no closed evolution
equation for either the first or the second moment of the momentum operator,
$\langle \mathsf{P} \rangle_{\rho_t}$ or $\langle \mathsf{P}^2
\rangle_{\rho_t}$, since due to the presence of the confluent hypergeometric
functions moments of arbitrary high order are involved in the equation.

However, the equations do get closed in the limit of a very massive test
particle close to thermal equilibrium, since in this case the velocity
$\tmmathbf{V}=\tmmathbf{P}/ M$ of the test particle is much smaller than the
typical velocity $v_{\beta}$ of the gas particles. Thus using $\tmmathbf{V}/
v_{\beta} \ll 1$ and $m / M \ll 1$ in (\ref{eq:p}) and (\ref{eq:e}) the
confluent hypergeometric functions are replaced by unity, since $\text{}_1 F_1
\left( \alpha, \gamma ; 0 \right) = 1$, and the reduced mass $m_{\ast}^{}$ is
replaced by $m$. This leads to
\begin{eqnarray}
  \frac{\mathd}{\mathd t} \langle \mathsf{P} \rangle_{\rho_t} & = & -
  n_{\tmop{gas}}  \frac{\sigma_{\tmop{tot}}}{4 \pi} \frac{32}{3 M} 
  \sqrt{\frac{2 \pi m}{\beta} } \langle \mathsf{P} \rangle_{\rho_t}
  \nonumber\\
  & = & - \eta \langle \mathsf{P} \rangle_{\rho_t} \nonumber
\end{eqnarray}
which describes velocity-proportional friction with coefficient $\eta$ leading
to the expected exponential relaxation to a mean momentum equal to zero.
Similarly,
\begin{eqnarray}
  \frac{\mathd}{\mathd t} \langle \mathsf{E} \rangle_{\rho_t} & = & - 2
  n_{\tmop{gas}}  \frac{\sigma_{\tmop{tot}}}{4 \pi} \frac{32}{3 M} 
  \sqrt{\frac{2 \pi m}{\beta} } \left( \langle \mathsf{E} \rangle_{\rho_t} -
  \frac{3}{2 \beta} \right) \nonumber\\
  & = & - 2 \eta \left( \langle \mathsf{E} \rangle_{\rho_t} - \frac{3}{2
  \beta} \right) \nonumber
\end{eqnarray}
shows that the mean kinetic energy relaxes exponentially to the equipartition
value ${3 / 2 k_{\text{B}} T}$. The relaxation rate
\begin{eqnarray}
  \eta & = & \frac{16}{3} n_{\tmop{gas}} \sigma_{\tmop{tot}}
  \sqrt{\frac{mk_{\text{B}} T}{2 \pi M^2} }, \nonumber
\end{eqnarray}
is equal to the result (\ref{eq:eta}) obtained in Sect. \ref{sec:Schrodinger}
for the considered case of a constant scattering cross-section.

\section{Conclusions}\label{sec:cc}

In conclusion, we discussed how the ``diffusive limit'' of the quantum
version of the linear Boltzmann equation yields the master equation for
quantum Brownian motion, and provides a microscopic formulation of the
relaxation and diffusion constants. In particular, we saw that this procedure
leads naturally to the{\tmem{ minimal extension}} required to turn the
Caldeira-Leggett master equation into Lindblad form. The approximations
invoked in this limit could be physically justified by using both the operator
and the Wigner-Weyl formulation of quantum mechanics in the Schr\"odinger
picture, while the Heisenberg picture provided a dynamic description of the
relaxation behaviour. Still, a mathematically more rigorous treatment of the
diffusive limit is clearly desirable, together with an extension of the
obtained results to the case of an arbitrary scattering cross-section.

\begin{acknowledgement}
The work was partially supported by the DFG Emmy Noether program (KH) and by
the Italian MIUR under PRIN05 (BV).
\end{acknowledgement}


\begin{thebibliography}{10}
  \bibitem[1]{Caldeira1983a}A.O. Caldeira, A.J. Leggett, Physica A
  \textbf{121}, 587 (1983)
  
  \bibitem[2]{Feynman1963a}R.P. Feynman, F.L. Vernon, Ann. Phys. (N.Y.)
  \textbf{24}, 118 (1963)
  
  \bibitem[3]{Haake1985a}F.~Haake, R.~Reibold, Phys. Rev.~A \textbf{32},
  2462 (1985)
  
  \bibitem[4]{Grabert1988a}H.~Grabert, P.~Schramm, G.~Ingold, Phys. Rev.
  \textbf{168}, 115 (1988)
  
  \bibitem[5]{Unruh1989a}W.G. Unruh, W.H. Zurek, Phys. Rev.~D \textbf{40},
  1071 (1989)
  
  \bibitem[6]{Hu1992a}B.L. Hu, J.P. Paz, Y.~Zhang, Phys. Rev.~D \textbf{45},
  2843 (1992)
  
  \bibitem[7]{Tameshtit1996a}A.~Tameshtit, J.~Sipe, Phys. Rev. Lett.
  \textbf{77}, 2600 (1996)
  
  \bibitem[8]{Hornberger2007LNP}See, e.g., K. Hornberger, Introduction to
  Decoherence Theory, eprint quant-ph/0612118
  
  \bibitem[9]{Breuer2002a}H.P. Breuer, F.~Petruccione, \textit{The Theory of
  Open Quantum Systems} (Oxford University Press, Oxford, 2002)
  
  \bibitem[10]{Sandulescu1987a}A.~S\u{a}ndulescu, H.~Scutaru, Ann. Phys. (N.Y.)
  \textbf{173}, 277 (1987)
  
  \bibitem[11]{Diosi1993a}L.~Di\'osi, Europhys. Lett. \textbf{22}, 1 (1993)
  
  \bibitem[12]{Diosi1995a}L.~Di\'osi, Europhys. Lett. \textbf{30}, 63 (1995)
  
  \bibitem[13]{Halliwell1995a}J.~Halliwell, A.~Zoupas, Phys. Rev. D
  \textbf{52}, 7294 (1995)
  
  \bibitem[14]{Barnett2005a}S.M. Barnett, J.D. Cresser, Phys. Rev.~A
  \textbf{72}, 022107 (2005)
  
  \bibitem[15]{Cercignani1975a}C.~Cercignani, \textit{Theory and application
  of the Boltzmann equation} (Scottisch Academic Press, Edinburgh, 1975)
  
  \bibitem[16]{Vacchini2000a}B.~Vacchini, Phys. Rev. Lett. \textbf{84}, 1374
  (2000)
  
  \bibitem[17]{Hornberger2006b}K.~Hornberger, Phys. Rev. Lett. \textbf{97},
  060601 (2006)
  
  \bibitem[18]{Joos1985a}E.~Joos, H.D. Zeh, Z. Phys. B: Condens. Matter
  \textbf{59}, 223 (1985)
  
  \bibitem[19]{Gallis1990a}M.R. Gallis, G.N. Fleming, Phys. Rev.~A
  \textbf{42}, 38 (1990)
  
  \bibitem[20]{Hornberger2003b}K.~Hornberger, J.E. Sipe, Phys. Rev.~A
  \textbf{68}, 012105 (2003)
  
  \bibitem[21]{Vacchini2001a}B.~Vacchini, Phys. Rev.~E \textbf{63}, 066115
  (2001)
  
  \bibitem[22]{Vacchini2001b}B.~Vacchini, J.~Math. Phys. \textbf{42}, 4291
  (2001)
  
  \bibitem[23]{Dodd2003a}P.J. Dodd, J.J. Halliwell, Phys. Rev.~D
  \textbf{67}, 105018 (2003)
  
  \bibitem[24]{Hornberger2007b}K.~Hornberger, EPL \textbf{77}, 50007 (2007)
  
  \bibitem[25]{Holevo1996a}A.S. Holevo, J.~Math. Phys. \textbf{37}, 1812
  (1996)
  
  \bibitem[26]{Petruccione2005a}F.~Petruccione, B.~Vacchini, Phys. Rev.~E
  \textbf{71}, 046134 (2005)
  
  \bibitem[27]{Vacchini2005a}B.~Vacchini, Phys. Rev. Lett. \textbf{95},
  230402 (2005)
  
  \bibitem[28]{Hornberger2004a}K.~Hornberger, J.E. Sipe, M.~Arndt, Phys.
  Rev.~A \textbf{70}, 053608 (2004)
  
  \bibitem[29]{Hornberger2003a}K.~Hornberger, S.~Uttenthaler, B.~Brezger,
  L.~Hackerm\"uller, M.~Arndt, A.~Zeilinger, Phys. Rev. Lett. \textbf{90},
  160401 (2003)
  
  \bibitem[30]{Hackermuller2003b}L.~Hackerm\"uller, K.~Hornberger, B.~Brezger,
  A.~Zeilinger, M.~Arndt, Appl. Phys.~B \textbf{77}, 781 (2003)
  
  \bibitem[31]{Arndt2005a}M.~Arndt, K.~Hornberger, A.~Zeilinger, Physics World
  \textbf{18}(3), 35 (2005)
  
  \bibitem[32]{Uhlenbeck1948a}C.S.W. Chang, G.E. Uhlenbeck, \textit{The
  kinetic theory of gases}, in \textit{Studies in statistical mechanics},
  edited by J.D. Boer (North-Holland, Amsterdam, 1970), Vol.~5
  
  \bibitem[33]{Schiff1968a}L.I. Schiff, \textit{Quantum Mechanics}, 3rd~edn.
  (McGraw-Hill, New York, 1968)
  
  \bibitem[34]{Adler2006a}S.L. Adler, J.~Phys.~A: Math. Gen. \textbf{39},
  14067 (2006)
  
  \bibitem[35]{Vacchini2002a}B.~Vacchini, Phys. Rev.~E \textbf{66}, 027107
  (2002)
  
  \bibitem[36]{Williams1966}M.M.R. Williams, \textit{The Slowing Down and
  Thermalization of Neutrons} (North-Holland, Amsterdam, 1966)
  
  \bibitem[37]{Wolfram}See e.g. http://functions.wolfram.com/
  
  \bibitem[38]{Gradshteyn1965a}I.S. Gradshteyn, I.M. Ryzhik, \textit{Table
  of integrals, series and products} (Academic Press, 1965)
\end{thebibliography}
\end{document}